\documentclass[a4paper]{jpconf}
\usepackage{graphicx}
\begin{document}
\title{Behaviour of the lateral shower age of cosmic ray extensive air
showers}
\author{Rajat K. Dey and Animesh Basak}
\address{Department of Physics, University of North Bengal, Siliguri, West Bengal, INDIA 734 013}
\ead{rkdey2007phy@rediffmail.com}

\begin{abstract}
Some simple arguments are introduced for a possible explanation of the
behavior of the lateral shower age of proton-initiated showers. The corresponding
analytical treatment based on the proposed argument is then illustrated. Using the
Monte Carlo simulation (MC) code CORSIKA, we have validated how the different
characteristics associated with the lateral shower age predicted in the present
analytical parametrization, can be understood. The lateral shower age of a proton-initiated
shower and its correlations with the lateral shower ages of electron- and
neutral pion-initiated showers supports the idea that the result of superposition of
several electromagnetic sub-showers initiated by neutral pions might produce the
lateral density distribution of electrons of a proton initiated shower. It is also noticed
with the simulated data that the stated feature still persists even in the local shower
age representation.
\end{abstract}
\section{Introduction}
Some recent works have revisited various conceptual issues on the lateral shower age ($s$) of extensive air showers (EAS) initiated by primary protons (p)/nuclei [1]. Basically the shape of lateral density distribution (LDD) of shower electrons (i.e. $\rm e^{\pm}$) is indicated by the shower age in the EM cascade theory, and is fairly valid also for p/nuclei initiated showers. The $s$ parameter is being estimated either from the reconstruction of EASs or from the radial variation of local shower age parameter (LAP) [1-2]. It was found from both the experimental, and simulation data that the NKG type lateral density function (LDF) or other type of modified NKG LDFs with a single $s$ is not adequate to describe the LDD of electrons initiated by p/nuclei/$\gamma$-ray primaries at all radial distances from the EAS core accurately [1]. In the present work, we have noticed the above behaviour in the LDDs of electrons for simulated showers initiated by $e^{\pm}$s and $\pi^{0}$s as well. This indicates that the $s$ parameter varies with the radial distance, which is in character the shower age at a point, and was labeled as the local age parameter (LAP) by Capdevielle and Gawin [2]. To assign a single age parameter to each shower, some works took the minimum value of the LAP from its variation against the radial distance while some other works took some sort of averaging of different LAPs obtained between the first minima and the second maxima [1]. This work discusses how the different observed properties associated with $s$ can be understood more precisely with some simple analytical arguments, and application of EAS simulations.
\section{Simple analytical method}
A proton/nuclei initiated shower is assumed to be a result of superposition of several partial electron-photon sub-cascades started mostly from the decay of first generation $\pi^{o}$s of the EAS in the atmosphere. The LDD  of electrons of a particular $e - \gamma$ sub-cascade (say, the i$^{th}$ sub-cascade) is believed to be described by the NKG type LDF with a shower age $s_{i}$. Usually the LDD of electrons of p/nuclei-initiated EASs can also be described by the NKG function but with a different shower age. Hence, the superposition principle applied to $e - \gamma$ sub-cascades in a proton/nuclei shower follows,
\begin{equation}
N_{e}C(s)X^{s-2}{(1+X)^{s-4.5}} = \sum_{i}{|N_{e_{i}}C(s_{i})X^{s_{i}-2}
		{(1+X)^{s_{i}-4.5}|}}
\end{equation} 
where $N_{e}$, being the electron size (i.e. $e^{\pm}$ together) of the resultant LDD of electrons and, $N_{e_{i}}$ represents the electron size for the LDD of the i$^{th}$ electron-photon sub-cascade with shower age $s_{i}$. Let $\acute{s}$ be the lateral shower age of an equivalent EM cascade initiated by primary e/$\gamma$ of the p/nuclei-initiated shower. The EM cascade also possesses the electron size $N_{e}$, and is assumed to be equal to the electron size of the p/nuclei shower. Dividing eq. (1) by ${N_{e}C(\acute{s})X^{\acute{s}-2}\\{(1+X)^{\acute{s}-4.5}}}$, being the LDF describing the LDD of electrons of an equivalent EM cascade, one may then get the following equation after taking logarithm from both sides,  	
\begin{equation}
s=\acute{s} - \frac{ln[C(s)/C(\acute{s})]-ln~\sum_{i}\alpha_{i}C(s_{i})/C(\acute{s})h^{\delta_{i}}}{ln(h)}
\end{equation}	
with $\alpha_{i}=N_{e_{i}}/N_{e}$, $h=X(1+X)$ with $X=r/r_{m}$ and $\delta_{i}=s_{i}-\acute{s}$.
	
It is expected that all the three lateral shower ages appear in eq. (2) should be different from each other but the functions $C(s)$, $C(\acute{s})$ and $C(s_{i})$ do differ very negligibly (analysis of simulated data yields; $C(s)\approx 2.969\times 10^{-5}$, $C(\acute{s})\approx 2.789\times 10^{-5}$ and $C(s_{i})\approx 2.731\times 10^{-5}$ ). Taking $C(s)\approx C(\acute{s})\approx C(s_{i})$, we can rewrite eq. (2) as
	
\begin{equation}
s\approx\acute{s} + \frac{ln~\sum_{i}\alpha_{i}h^{\delta_{i}}}{ln(h)}
\end{equation}
	
Taking $N_{{e}_{i}}\approx n_{e}$ with $i=1,2,3,....n$), and $s_{i}\approx {\tilde{s}}$ for all sub-cascades. Let $\delta_{i}$ accounts the difference between the lateral shower ages of two EM cascades, in which one refers to the i$^{th}$ $e - \gamma$ sub-cascade initiated by a $\pi^{0}$, and the rest is the effective EM cascade generated by primary e/$\gamma$. It is further assumed that $\delta_{i}$ will not change appreciably with the atmospheric depth, and the ratio ${N_{{e}_{i}}}/N_{e}$ will similarly not vary from one electron-photon sub-cascade to the other. Under that circumstances we have obtained the following from eq. (2) 
\begin{equation}
s\approx\acute{s} - \delta = 2\acute{s}-\tilde{s} 
\end{equation}
with $\delta > 0$. One can rewrite eq. (2) in an alternative way, in terms of lateral density of shower electrons:
\begin{equation}
(\acute{s}-\tilde{s})\approx \frac{ln(n\alpha_{e}) - ln{[\frac{\rho_{Had}(r)}{\rho_{EM}(r)}]}}{ln[h]}\approx \delta
\end{equation}

$\delta$ can be estimated from the above relation by using simulations. The LAP of a p/nuclei-initiated shower is defined by [1]
\begin{equation}
s_{local}^{Had}(i,j)=\frac{ln(F_{ij}X_{ij}^{2}Y_{ij}^{4.5})}{ln(X_{ij}Y_{ij})} 
\end{equation}

For the equivalent EM cascade, the corresponding LAP is,

\begin{equation}
s_{local}^{EM}(i,j)=\frac{ln(\acute{F}_{ij}\acute{X}_{ij}^{2}\acute{Y}_{ij}^{4.5})}{ln(\acute{X}_{ij}\acute{Y}_{ij})} 
\end{equation}
The superposition principle yields,
\begin{equation}
s_{local}^{Had}(i,j)\approx {\frac{ln[(\sum_{k}\tilde{\rho}_{ij,k})\tilde{X}_{ij}^{2}\tilde{Y}_{ij}^{4.5}]}{ln(\tilde{X}_{ij}\tilde{Y}_{ij})}}
\end{equation}

The minimum LAP obtained from the LAP versus $r$ curve is used as a lateral shower age of a shower. In terms of minimum LAP, we have obtained; $s_{local}^{Had}(min)-\acute{s}_{local}^{EM}(min)\approx \delta$ and also the $\acute{s}_{local}^{EM}(min)-\tilde{s}_{local}(min)\approx \delta$. 

The simulated LDD data of electrons at different radial distances are fitted usually by the NKG type function for obtaining the parameter $s$. The longitudinal shape/age parameter ($s_{\parallel}$) of the energy and angle distributions of shower electrons of very high-energy to UHE MC showers is well approximated to a simple form as

\begin{equation}
s_{\parallel} = \frac{3 X} { X + 2 X_{\rm max}},
\end{equation}

where $X_{\rm max}$ is the depth of shower maximum and $X$ is the atmospheric slant depth of the KASCADE level (in unit of radiation lengths). Knowing both the depth parameters, the $s_{\parallel}$  of a shower can be estimated. Hence, one should use $s_{\parallel}$ instead of the depth $\rm X$ referring to the shower maximum. The value of $\rm X/X_{max}$ is the key for the description of the EAS development stage. This work however studies the difference between $s_{\parallel}$ and $s$ for p, $e^{-}$ and required number of $\pi^{0}$ generated simulated  showers.
 
\section{Results and discussions}
In the MC code CORSIKA ver. 7.4 [3], the high-energy EPOS-LHC [4] and low-energy UrQMD [5] models are combined for generating p, e/$\gamma$ and  $\pi^{0}$ showers at $E=2$~PeV. The EGS4 [6] program library deals with the EM interactions in an EAS development. FIXCHI$\approx 75$~gcm$^{-2}$ is used for $\pi^{0}$ showers to deliver better results.

The simulated densities for secondary electrons at different radial distances from the shower core ($10 - 225$~m) are estimated in the analysis for simulated p, an equivalent average EM cascade initiated by $e^{-}$ and the required $\pi^{0}$ showers. In the Fig. 1, we have plotted the mean electron densities over the above radial distance range for these simulated showers. For the stated three types of showers, a very good agreement among the LDDs of shower electrons is noted over the radial distance range $10 - 100$~m. It ensures the fact that the average LDD of electrons of p-initiated showers can be well represented by the superposition of a justifiable number of electron-photon sub-cascades from the decay of secondary $\pi^{0}$s produced after the first interaction point of the same p shower.
\begin{figure}[h]
\begin{minipage}{14pc}
\includegraphics[trim=-0.6cm 0.6cm 0.6cm 1.05cm, clip=true, totalheight=0.22\textheight, angle=0]{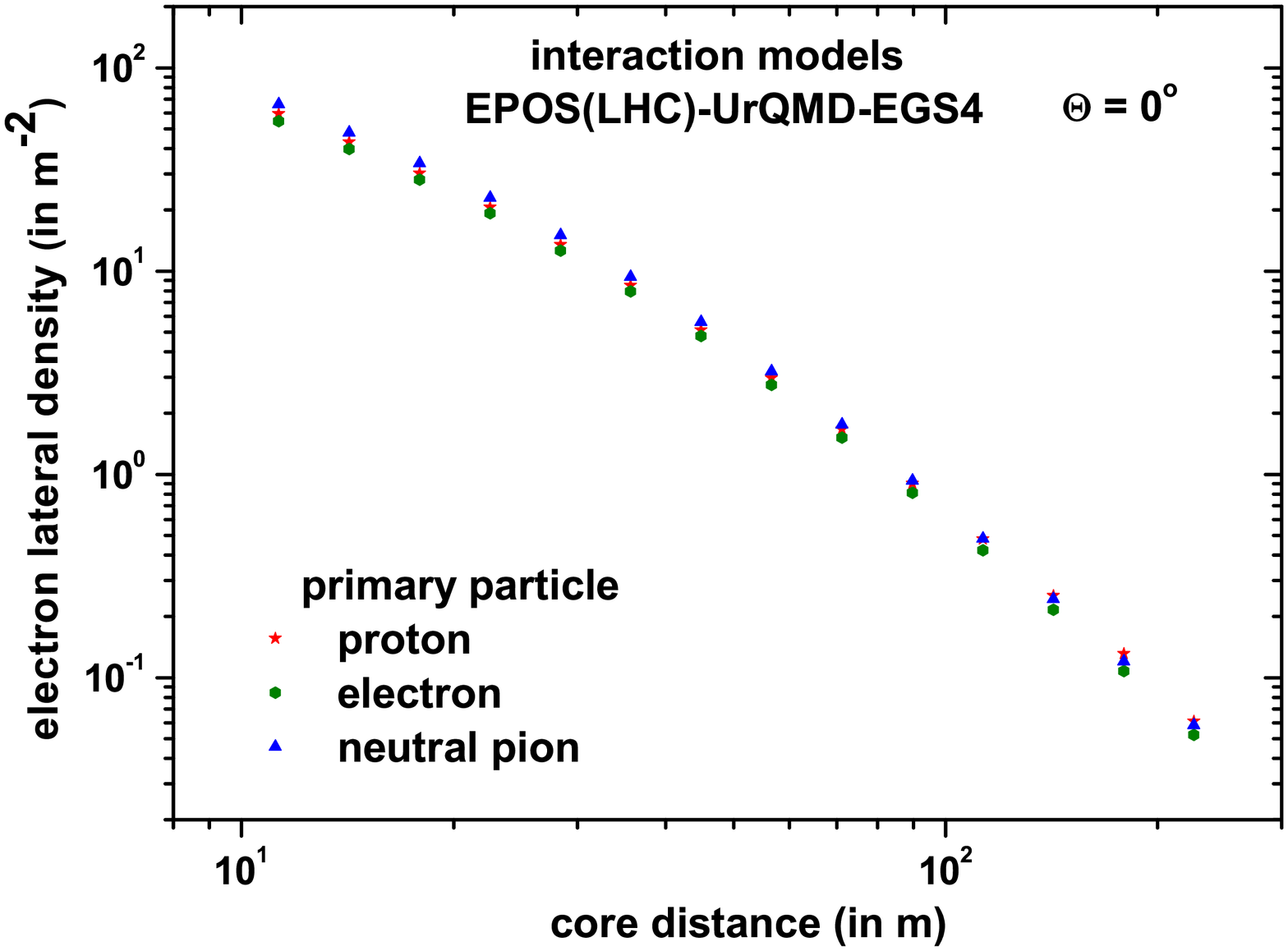}
\caption{\label{label}Variation of $\rho_{e}$ with $r$.}
\end{minipage}\hspace{5pc}%
\begin{minipage}{15pc}
\includegraphics[trim=-.6cm 0.6cm 2.8cm 1.05cm, clip=true, totalheight=0.22\textheight, angle=0]{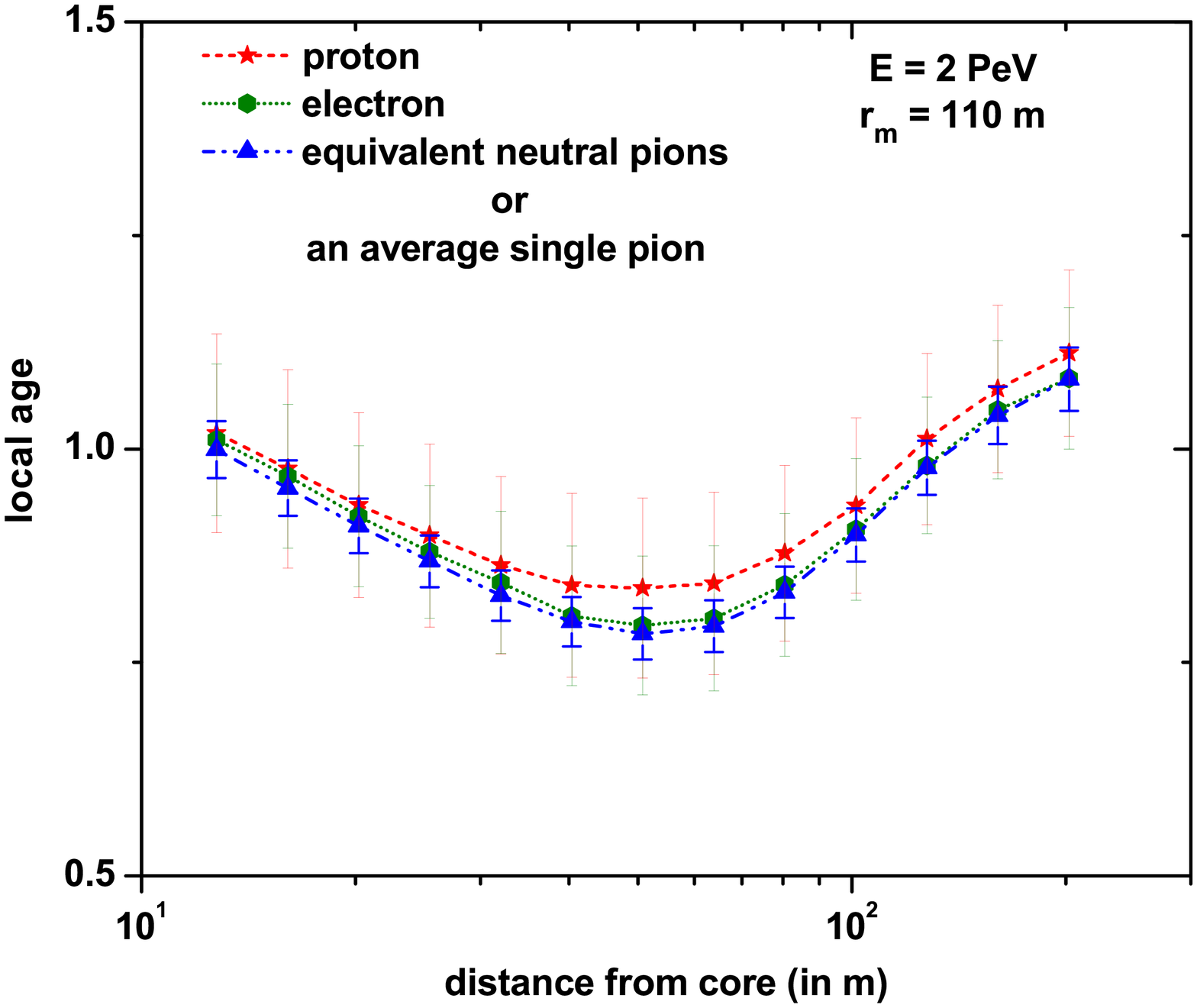}
\caption{\label{label}Variation of LAP with $r$.}
\end{minipage}
\end{figure}
Analysis of data results, $\delta\approx s-\acute{s}=0.053$ and $2\acute{s}-\tilde{s}=0.899\approx s$. When MC data are used in eq. (6), we have obtained, $\delta\approx \acute{s}-\tilde{s}\approx 0.05$. 
\begin{figure}[h]
\begin{minipage}{12pc}
\includegraphics[trim=0.6cm 0.6cm 0.6cm 1.05cm, clip=true, totalheight=0.15\textheight]{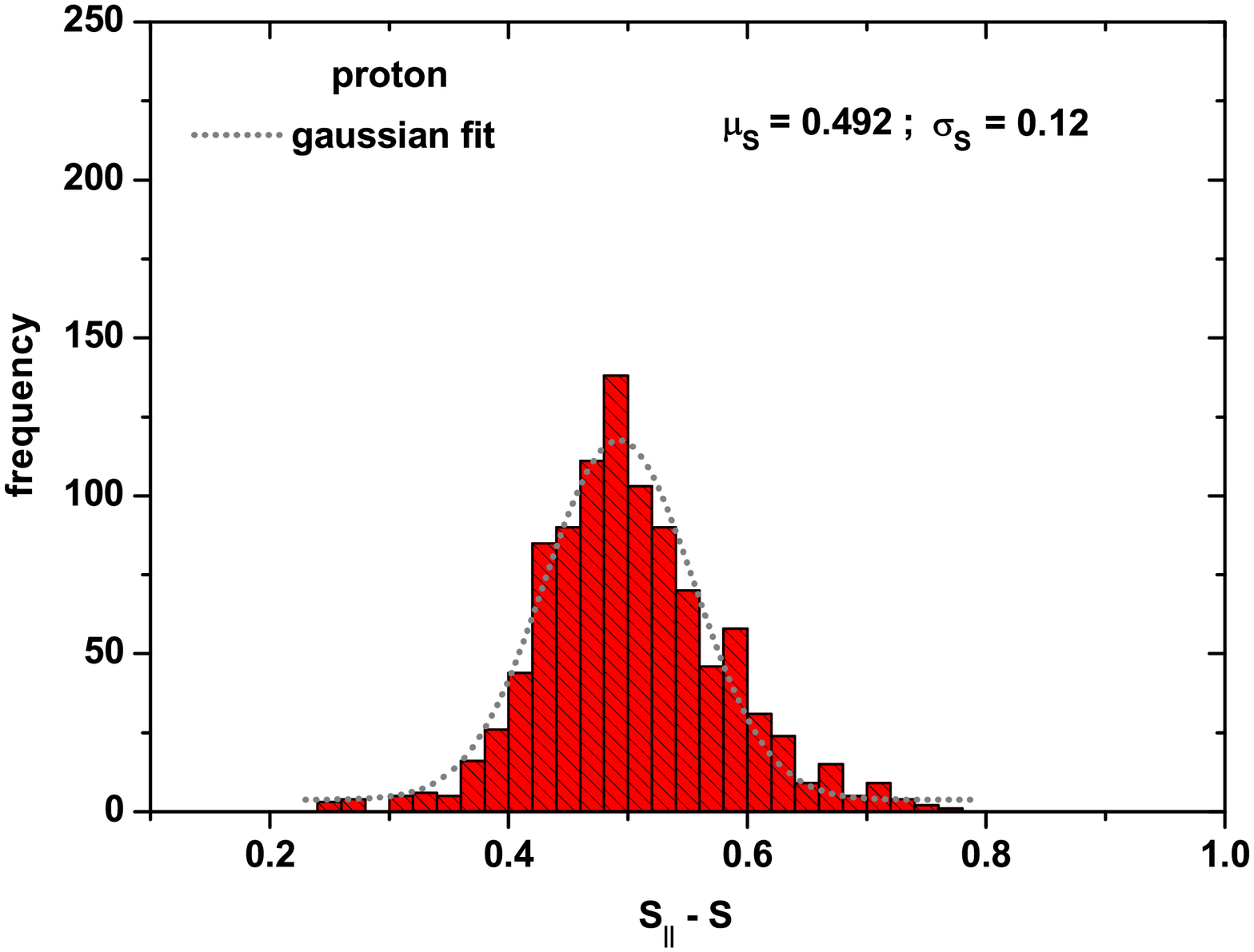}
\caption{\label{label}Frequency distribution of $(s_{\parallel}-s)$ for p showers.}
\end{minipage}\hspace{1pc}%
\begin{minipage}{12pc}
\includegraphics[trim=0.6cm 0.6cm 0.6cm 1.05cm, clip=true, totalheight=0.15\textheight]{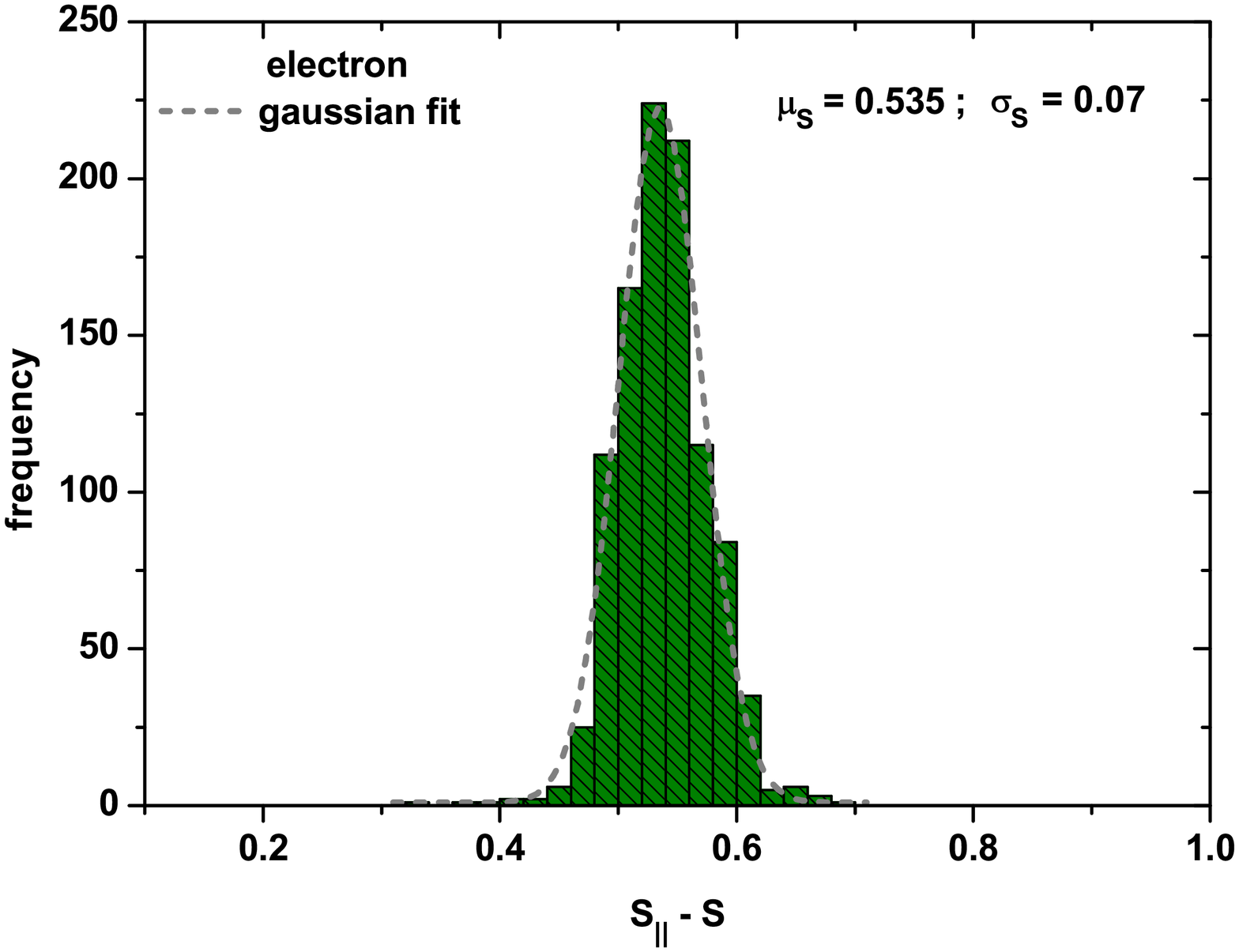}
\caption{\label{label}Frequency distribution of $(s_{\parallel}-s)$ for $e^{-}$ showers.}
\end{minipage}\hspace{1pc}%
\begin{minipage}{12pc}
\includegraphics[trim=0.6cm 0.6cm 0.6cm 1.05cm, clip=true, totalheight=0.15\textheight]{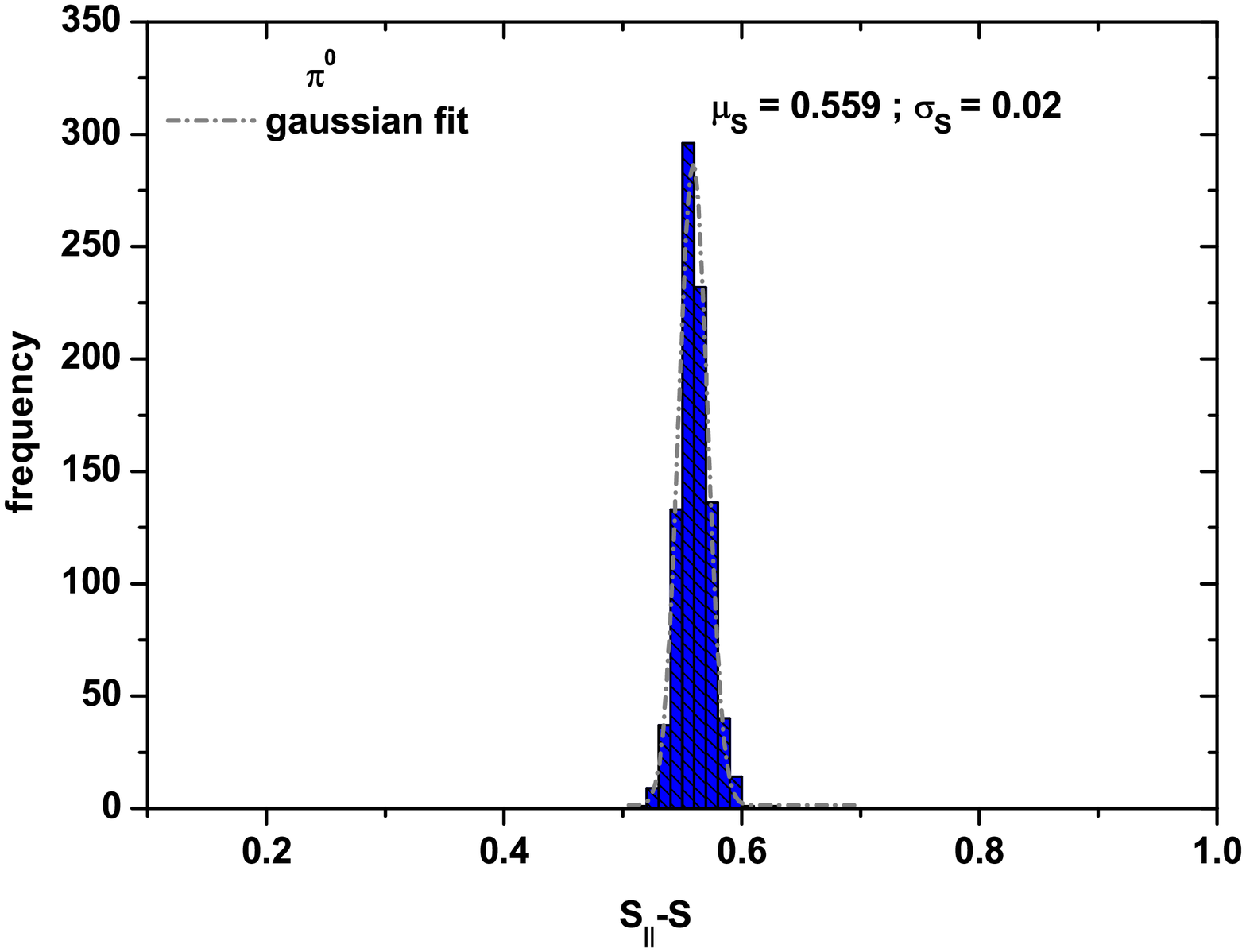}
\caption{\label{label}Frequency distribution of $(s_{\parallel}-s)$ for $\pi^{0}$ showers.}
\end{minipage} 
\end{figure}
The present analytical approach may receive more impetus if we compute the parameter $\delta$ in terms of the LAPs introduced in Sec. 2 for simulated p-, $e^{-}$- and $\pi^{0}$-initiated showers. The variation of LAP versus $r$ is shown in Fig. 2. The error of the LAP is found $\approx 0.04$ for $12<r<205~$m. The minimum LAP from LAP versus $r$ variation at about $50~$m is taken as the lateral shower age of an EAS. We have obtained $\delta\approx s_{local}(min)-\acute{s}_{local}(min)\approx 0.044$ and $\delta \approx \acute{s}_{local}(min)-\tilde{s}_{local}(min)\approx 0.01$. We have noticed a rise in $\delta$ with $r$.

The present MC simulation gives the frequency distribution peaks within the range $\approx 0.50 - 0.55$, which is consistent with some early MC simulation predictions satisfying a general relation of the form, $s_{\parallel} - s\approx 0.5$  or equivalently $s_{\parallel} \sim 1.4{s}$. The correlations between $s_{\parallel}$ and $s$ reveal that an average p shower at least can be explained as the result of superposition of several EM sub-showers generated by $\pi^{0}s$.    
\section{Conclusions}
The simulated electron LDDs of p-, $e^{-}$- and $\pi^{0}$-initiated showers unequivocally support the idea, explained in the adopted simple analytical argument. The difference $s_{\parallel}-s$ can be explained as a result of superposition of EM sub-showers initiated by $\pi^{0}$s, and are found sensitive to CR mass composition. The numerical value of the difference $\delta\approx s-\acute{s}=0.053\neq 0$ indicates that the radial dependence of $s$ is different than that of $\acute{s}$. This otherwise supports the present view of the consideration of the superposition principle that indeed could explain the behavior of the shape parameter of a p/nuclei initiated shower. The value of $\delta$ is almost recovered in the language of LAP i.e. $s_{local}(min)-\acute{s}_{local}(min)$ but the value $\acute{s}_{local}(min)-\tilde{s}_{local}(min)$ deviates much from its earlier value in terms of $s$ obtained from shower reconstruction.
\section*{Acknowledgments}
\noindent    
RKD acknowledges the financial support under grant. no. 1513/R-2020 from the University of North Bengal.

\end{document}